\documentstyle[12pt]{article}
\newtheorem{verticalk}{Definition}
\newtheorem{vertical0}[verticalk]{Definition}
\newtheorem{cft}[verticalk]{Definition}
\newtheorem{symplectic}[verticalk]{Definition}
\newtheorem{vertical2}[verticalk]{Definition}
\newtheorem{HL}[verticalk]{Definition}
\newtheorem{pft}[verticalk]{Definition}
\newtheorem{Louvill}[verticalk]{Definition}
\newtheorem{Hodge}[verticalk]{Definition}
\newtheorem{bigraded}[verticalk]{Definition}
\newtheorem{Dirac}[verticalk]{Definition}
\newtheorem{1}{Lemma}
\newtheorem{HL-type}[1]{Lemma}
\newtheorem{annul}[1]{Lemma}
\newtheorem{wHodge}[1]{Lemma}
\newtheorem{mechanics}{Proposition}
\newtheorem{Gawedzki}{Corollary}
\newtheorem{C1}[Gawedzki]{Corollary}
\newtheorem{seven}[Gawedzki]{Corollary}
\begin{document}
\title{CLASSICAL FIELD THEORY\\AND ANALOGY BETWEEN\\NEWTON'S AND MAXWELL'S
EQUATIONS}
\author{ZBIGNIEW OZIEWICZ\thanks{On leave of absence from University of
Wroc{\l}aw, Poland. Partially
supported by Polish Committee for
Scientific Research, Grant 2 2419 92 03/92.}\\
Universidad Nacional Autonoma de M\'exico\\oziewicz@redvax1.dgsca.unam.mx}
\date{October 1993}
\maketitle
\newcommand{\ds}{\displaystyle}
\newcommand{\un}{\underline}
\newcommand{\be}{\begin{equation}}\newcommand{\ee}{\end{equation}}
\newcommand{\ba}{\begin{array}}\newcommand{\ea}{\end{array}}
\newcommand{\bd}{\begin{description}}\newcommand{\ed}{\end{description}}
\newcommand{\ra}{\longrightarrow}
\newcommand{\la}{\longleftarrow}
\newcommand{\Ra}{\Longrightarrow}
\newcommand{\LRa}{\Longleftrightarrow}
\newcommand{\hr}{\hookrightarrow}
\newcommand{\hl}{\hookleftarrow}
\newcommand{\id}{\mbox{id}}
\newcommand{\der}{\mbox{der}}
\newcommand{\ring}{\mbox{ring}}
\newcommand{\mod}{\mbox{mod}}
\newcommand{\alg}{\mbox{alg}}
\newcommand{\image}{\mbox{image}}
\newcommand{\End}{\mbox{End}}
\newcommand{\gen}{\mbox{gen}}
\newcommand{\lin}{\mbox{lin}}
\newcommand{\vol}{\mbox{vol}}
\newcommand{\codim}{\mbox{codim}}
\newcommand{\Ver}{\mbox{Ver}}
\newcommand{\Hor}{\mbox{Hor}}
\newcommand{\cVer}{{\cal V}er}
\newcommand{\grade}{\mbox{grade}}
\newcommand{\eVer}{\em Ver}
\newcommand{\eHor}{\em Hor}
\newcommand{\egrade}{{\em grade}}
\newcommand{\ebigrade}{\em bigrade}
\newcommand{\eid}{{\em id}}
\newcommand{\eder}{{\em der}}
\newcommand{\ering}{{\em ring}}
\newcommand{\emod}{{\em mod}}
\newcommand{\ealg}{{\em alg}}
\newcommand{\eimage}{{\em image}}
\newcommand{\eEnd}{{\em End}}
\newcommand{\egen}{{\em gen}}
\newcommand{\elin}{{\em lin}}
\newcommand{\edeg}{{\em deg}}
\newcommand{\edim}{{\em dim}}
\def\choose#1#2{\left(\begin{array}{c}{{#1}}\\{{#2}}\end{array}\right)}
\def\ss{\begin{tabular}{c}}
\def\kk{\end{tabular}}
\def\Zint{{Z\kern -.45em Z}}
\def\Nint{{I\!\!N}}\def\Rint{{I\!\!R}}
\begin{abstract} A bivertical classical field theory
include the Newtonian mechanics and Maxwell's electromagnetic field
theory as the special cases. This unification allows to recognize the formal
ana
Newtonian mechanics and Maxwell's electrodyna\-mics.
\end{abstract}
\begin{flushright}
Dedicated to Professor Constantin Piron
\end{flushright}
\newpage
\subsection*{Contents}
\bd
\item{} Foreword
\item{I.} Axiomatique classique
\begin{enumerate}
\item Vertical distribution and filtration of forms
\item Classical field theory
\item Submanifolds
\item Hamilton-Lagrange field theory
\end{enumerate}
\item{II.} Phenomenology
\begin{enumerate}
\item Phenomenological field theory
\item Louville's differential forms
\item Poincar\'e-Cartan submanifolds
\end{enumerate}
\item{III.} Field equations
\begin{enumerate}
\item The Hodge's map
\item Calculus with the splitting
\item Dirac operator and codifferential
\item Legendre's transforms
\item Second order field equations
\end{enumerate}
\item{} References
\ed

\subsection*{Foreword}
This paper is inspired by Professor Constantin Piron.\footnote{Professor Piron
visited University of Wroc{\l}aw the first time in 1980.
During Martial
Law in Poland, when I was in prison in 1982, and again in 1984, Professor Piron
bravely keep scientific contact with me, supporting my family and Solidarity,
which required courage and a good heart.}.
Professor Piron's credo is the
unity of human's knowledge gained by mathematicians, physicists, philosophers,
theologians. Piron is known, among other,
as the creator of the {\em New Quantum Mechanics} (1983), a {\em realistic}
theo
{\em unifying} the classical and quantum physics.

The main ingredients of the classical physics are Newtonian mecha\-nics and
Maxwell's electrodynamics. These phenomenological theories are related to
the diverse phenomena and are presented as disconnected in a
separate courses, textbooks and monographs. The one aim of this paper is to
exibits the formal analogies among these theories in the framework of the {\em
almost} multisymplectic geometry, $d\Omega\neq 0.$

We consider the bundle of the exterior differential forms, in mechanics of
dimension $1+4n,$ in electromagnetism of dimension $4+20n,$ with the
phenomenological {\em not} presymplectic differential form. Our theory is
bivertical (definition 5) for arbitrary dimensional space-time manifold.
We determine the
{\em four} Poincar\'e-Cartan subbundles: hamiltonian, lagrangian and two new
not-named subbundles. This leads to the {\em twelve} Legendre's transforms
among
these subbundles, of which {\em two} are the well known.

The field equations of considered bivertical theory for $n=1$ reduce to that of
Newton's equations and for n=4 to Maxwell's equations. This unification allows
to see analogies among the notions of Newtonian dynamics and of Maxwell's
theory. In particular, {\em force field} $\leftrightarrow$ {\em current}, the
London's equation in electromagnetism is an analogy of the harmonic oscilator
force in Newton's dynamics, one can pose the Kepler's problem in Maxwell's
electrodynamics by formal analogy to the Kepler's problem in mechanics, etc.
In the non-linear model we discuss the second order field equations.

The limited space do not allows to include here the discussion of the
conservation laws, Noether currents, energy-momentum tensor, Poynting
differential form etc., from the point of view of the presented formal
unification.

It is a pleasure to thank Professor Constantin Piron for many inspiring
discussion during the last 14 years of our friendship. The subject of the paper
was presented in the fall 1992 at the University of Wroc{\l}aw.
Some aspects related to this paper are elaborated by Magdalena Gusiew in the
diploma thesis (1993). I am gratefull to Magdalena Gusiew and Grzegorz
Jastrz{\c
e}bski for iluminating conversations.

\vspace{0.1cm}\noindent{\bf History.} Multisymplectic geometry in classical
field theory was initiated by Dedecker in
1953 and was developed in Warsaw by W{\l}odzimierz Tulczyjew around 1968, and
by Kijowski (1973), Gaw{\c e}dzki (1972), Szczyrba and Kondracki (1979). In
Chechia by Krupka since 1975. See Kijowski and Tulczyjew (1979).\\

\vspace{0.1cm}\noindent{\bf Notations.}\\
$\Lambda\equiv\Lambda_E\equiv\oplus\Lambda^k$ is de Rham complex of the
differential forms on a manifold $E$, ${\cal F}\equiv\Lambda^0,$
$Z\equiv\{\alpha\in\Lambda,\,d\alpha=0\}.$\\
$W\equiv W_E\equiv\der{\cal F}$ is the Lie ${\cal F}$-module of the ({\bf
one})-
such that $W^\wedge\equiv\oplus W^{\wedge k}$ is the Grassmann ${\cal
F}$-algebra and graded Lie $I\!\!R$-algebra of multivector fields, $W^{\wedge
0}
The left Grassmann multiplication (an algebra map), $e:\Lambda\ra\lin\Lambda,$
is
$e_\alpha\beta\equiv\alpha\wedge\beta$, $e$ for {\em exterior}. Then $i$ denote
the interior product, $i:W^\wedge\ra\lin\Lambda,$ which is anti-algebra map.\\
$|\alpha|\equiv\grade\,\alpha\in\Nint$ and
$\psi$ denote the automorphism of Grassmann algebras,
$\psi\alpha\equiv(-1)^{|\alpha|}\alpha.$

\section*{I. Axiomatique classique}
\subsection*{Vertical distribution and filtration of forms}
Let $E$ be a manifold with the distribution Ver$\,\subset W$ which is said to
be
the {\em vertical} distribution.
\begin{verticalk}
The {\em\cal F}-submodule
$$\Lambda_{(k)}\equiv\{\alpha\in\Lambda,\,i_Z\alpha=0\quad\forall\:Z\in\eVer^{\w
is said to be the submodule of $k$-{\em vertical} forms. The factor module is
denoted by
$\Lambda_{[k]}\equiv\Lambda_{(k)}/\Lambda_{(k-1)}$ and if $\alpha\in\Lambda$
then $\alpha/(k)\in\Lambda/\Lambda_{(k)}.$
\end{verticalk}
\begin{C1}
$\Lambda_{(k)}\wedge\Lambda_{(l)}\subset\Lambda_{(k+l)},\;
\Lambda_{[k]}\wedge\Lambda_{[l]}\subset\Lambda_{[k+l]}$ and we have the {\em
filtration} of forms
$$\Lambda_{(0)}\subset\ldots\subset\Lambda_{(k)}\subset\Lambda_{(k+1)}\subset\ld
\end{C1}
Let $\cVer$ be the submodule of the differential one-forms
anihilating Ver
$$\cVer\equiv\{\alpha\in\Lambda^1;\;\alpha(\Ver)=0\}.$$
\begin{vertical0} The differential form $\alpha$ on $\{E,\eVer\}$ is said to be
{\em vertical} if $\alpha\in\cVer^\wedge.$ $\cVer^{\wedge 0}\equiv{\cal F}_E.$
\end{vertical0}
Let for $\alpha\in\Lambda,$ Dist$\,\alpha$ be the associative
distribution of $\alpha,$
$$\mbox{Dist}\,\alpha\equiv\{X\in W;\;i_X\alpha=0\}.$$
The form $\alpha$ is decomposable iff
$$\dim E=|\alpha|+\dim\,\mbox{Dist}\alpha.$$
\begin{1}.\\
\bd
\item{(i)} The form $\alpha$ is $0$-vertical iff $\alpha$ is a
vertical, $\Lambda_{(0)}=\cVer^\wedge,$
$$i_{\eVer}\alpha=0\quad\LRa\quad \alpha\in \cVer^{\wedge}.$$
\item{(ii)}$$\left\{\ba{l}\theta\in\cVer^\wedge\\|\theta|=\dim\,\cVer\ea\right\}
\left\{\ba{l}{\em Dist}\,\theta=\eVer\\\theta\;{\em is\,
decomposable}\ea\right\
\ed\end{1}
There is the one to one correspondence between the (vertical) distributions
and
one-dimensional modules of decomposable forms. We will identify
$$\{E,\theta\}\equiv\{E,\Ver\equiv\mbox{Dist}\,\theta\},$$
where the {\em decomposable} form $\theta$ is defined up to the multiplication
b
${\cal F}_E.$ The distribution Ver is locally integrable iff
$d\theta$ is 1-vertical and then $\{E,\theta\}$ is locally fibered,
$d\theta\in\Lambda_{(1)}\LRa d(f\theta)\in\Lambda_{(1)}$ for $0\neq
f\in{\cal F}.$

If $E$ is fibered over oriented manifold $\{M,\vol\},$ $E\stackrel{\pi}{\ra}M,$
then $\theta\equiv\pi^*\vol$ is a decomposable cocycle.

\newpage
\subsection*{Classical field theory}
\begin{cft}.\begin{description}
\item{(i)} The classical field theory is a triple $\{E,\eVer,\Omega\}$, where
$E
manifold, {\eVer} is a distribution and $\Omega$ is a differential form on
$E$ such that $|\Omega|=1+{\em codim}{\eVer}>1.$
\item{(ii)} The submanifold $\phi$ of $E$ is said to be the solution (the
space-time ma\-ni\-fold)
of $\{E,{\em Dist}\theta\equiv\eVer,\Omega\}$ if for every vector field $Z$ on
$
\be \phi^*\theta\neq 0\quad\mbox{and}\quad\phi^*i_Z\Omega=0.\ee
\item{(iii)} The field theory $\{E,\eVer,\Omega\}$ is said to be {\em regular}
i
integrable distribution {\eHor} tangent to the solutions of the the equations
(1
is complementary to {\eVer},
$${\eHor}\cap{\eVer}=0\quad\mbox{and}\quad W={\eHor}\cup{\eVer}.$$
\end{description}\end{cft}
{\bf Comment.} The field theory is
regular if every solution $\phi$ (1) is transversal to Ver and
dim$\phi=\codim\,\Ver$ $\equiv|\theta|$.

The form $\Omega$ determine the $\cal F$-linear map\footnote{The form
$\Omega$ can be viewed as the retrangular matrix
${\choose{\dim E}{|\theta|}} \times(\dim E).$}
$$\Omega:\;W^{\wedge|\theta|}\ra\Lambda^1.$$
\begin{mechanics} Let $\ker\Omega\subset W^{\wedge|\theta|}.$ Then
\begin{description}
\item{(i)}\hspace{2.0cm} $\dim\ker\Omega=1\quad\Ra\quad\{E,\eVer,\Omega\}$ is
re
\item{(ii)} Let $\{E,\eVer,\Omega\}$ be regular, {\em
codim}$\eVer\equiv|\theta|
$\Omega$ be a cocycle (so $\Omega$ is symplectic).
Then $\dim\ker\Omega=1,$ ($\Ra\;\dim E=$odd).
\end{description}
\end{mechanics}
{\bf Comment.} If $|\theta|=1,$ then a cocycle $\Omega$ is regular iff
$\dim\ker\Omega=1.$ The $|\theta|=1$ refers to mechanics and the property
to be regular is said to be the
{\em classical determinism}\footnote{The symplectic mechanics, $d\Omega=0$ and
$\dim\ker\Omega=1,$ on jet manifolds of arbitrary order has been presented by
Olga Krupkova (1992).}.

\vspace{0.1cm}\noindent{\bf Example.} Regular field theory
$\{E,\theta,\Omega\}$
need not imply that dim$\,\ker\Omega=1.$ Let dim$\,E=1+4n$ with a chart
$\{t,q^A,v^A,p_A,f_A\}$ and
$\theta\equiv dt.$ Let $\Omega\equiv (dp_A-f_Adt)\wedge(dq^A-v^Adt),$ then
$d\Omega\neq 0,$ $\dim\ker\Omega=1+2n$ and this mechanics $\{E,\theta,\Omega\}$
is {\em regular}.

\vspace{0.2cm}\noindent{\bf Proof of Proposition 1.} The integrable
distribution
Hor$\,\subset W$ tangent to the solutions of the field equations (1) needs to
sa
the two conditions
\begin{itemize}
\item $\theta(\Hor^{\wedge|\theta|})\neq 0\qquad(\Ra\quad\dim
\Hor\geq|\theta|+\dim(\Hor\cap\Ver)),$
\item $\Hor^{\wedge|\theta|}\subset\ker\Omega
,\quad\left(\Ra\quad\dim(\Hor^{\wedge|\theta|})=\choose{\dim
\Hor}{|\theta|}\leq\dim\ker\Omega\right).$
\end{itemize}
The last condition imply
$$\dim\ker\Omega=1\quad\Ra\quad\dim \Hor\leq|\theta|.$$
It follows that Hor$\,\cap$Ver$=0$ and dim$\,\Hor=|\theta|\equiv\codim\,\Ver,$
which complete the proof of (i).

Let $\Omega$ be a cocycle. Then the associated distribution, $\ker\Omega\subset
is integrable. If $|\theta|=1,$ then Hor$=\ker\Omega$ is integrable and the
regularity of $\Omega$ imply that dim$\,\ker\Omega=1.$
\hfill$\Box$\\
\begin{Gawedzki}[Gaw{\c e}dzki 1972]
Let $\{E,\eVer,\Omega\}$ be regular field theory. Then it is sufficient to
consider the field equations (1) for vertical vector fields only,
$$\phi^*i_{\eVer}\Omega=0\quad\Ra\quad\phi^*i_{W}\Omega=0 .$$
\end{Gawedzki}
{\bf Proof.} Let the distribution $\Hor$ be as in the proof of the Proposition
1
We need to show that $\Omega(\Ver\wedge \Hor^{\wedge|\theta|})=0\;\Ra\;
\Omega(W\wedge \Hor^{\wedge|\theta|})=0.$ This is the case if $W=\Hor\cup
\Ver$ and $\Hor^{\wedge(1+|\theta|)}=0.$
\hfill$\Box$\\

\subsection*{Submanifolds}
Let $0\leq i\leq|\Omega|$ and $0\leq j\leq 2.$
Let $\{\Psi_{3i+j}\hr\Psi_{3i-2}\hr E\}$ and
$\{\Omega_i\in\Lambda(\Psi_{3i-2}),\,\Omega_0\equiv\Omega\}$ be collections
of {\em maximal} submanifolds and differential forms {\em defined} by the
conditions

$\Psi^\ast\theta\neq 0,$\hfill
$(\Psi_{3i})^\ast d\Omega_i= 0,$\hfill
 $(\Psi_{3i+1})^\ast\Omega_i= d\Omega_{i+1},$\hfill
 $(\Psi_{3i+2})^\ast\Omega_i= 0.$

Therefore $\dim\Psi_{3i-1}=\dim\Psi_{3i}=\dim\Psi_{3i+1}$ and
$$\phi\hr\dots\hr\Psi_4\hr({\cal P}\equiv\Psi_1)\hr\Psi\hr E.$$
\begin{symplectic}.
\bd
\item{(i)} The submanifold $\Psi\hr E$ is said to be the {\em pre-symplectic}
manifold of $\{E,\theta,\Omega\}$ if $\Psi$ is a {\em maximal} submanifold
anihilating $d\Omega,$
$$
\Psi^*\theta\neq0\quad\mbox{and}\quad\Psi^*d\Omega=0.$$ The presymplectic
manifold $\Psi$ is said to be {\em symplectic} if the field theory
$\{\Psi,\Psi^*\theta,\Psi^*\Omega\}$ is regular. The regular cocycle
$\Psi^*\Omega$ is said to be the {\em symplectic} form on
$\{\Psi,\Psi^*\theta\}.$
\item{(ii)} The submanifold ${\cal P}\hr\Psi\hr E$ is said to be
{\em Poincar\'e-Cartan subma\-ni\-fold} (exact presymplectic) if $\cal P$ is a
{
maximal} submanifold on which $\Omega$ is exact,
$${\cal P}^*\theta\neq0\quad\mbox{and}\quad{\cal P}^*\Omega=d\alpha.$$
The presymplectic potential $\alpha\equiv\Omega_1$ is said to be the
Poincar\'e-Cartan form. If the field theory $\{{\cal P},{\cal
P}^*\theta,d\alpha
regular then $\alpha$ is said to be the {\em regular} Poincar\'e-Cartan form.
\item{(iii)} The submanifold ${\cal L}\equiv\Psi_2\hr\Psi\hr E$ is said to be
{\
manifold of $\{E,\theta,\Omega\}$ if $\cal L$ is a {\em maximal} submanifold
anihilating $\Omega,$
$${\cal L}^*\theta\neq0\quad\mbox{and}\quad{\cal L}^*\Omega=0.$$
\item{(iv)} The submanifold $\Psi_4\hr{\cal P}\hr\Psi\hr E$ is said to be the
{\em Hamilton-Jacobi} manifold if $\Psi_4$ is a {\em maximal} submanifold on
which the Poincar\'e-Cartan differential form is exact,
$${\Psi_4}^*\theta\neq0\quad\mbox{and}\quad{\Psi_4}^*\alpha=d\Omega_2.$$
The potential $\Omega_2$ is said to be the Hamilton-Jacobi differential form,
$|\Omega_2|=|\theta|-1.$ The equations in (iii-iv) are said to be the
Hamilton-J
equations.
\ed\end{symplectic}

On presymplectic submanifold $\Psi^\ast\Omega$ is a cocycle, the action
integral
Oziewicz 1992) and the field equations of the definition 3 are the
Euler-Lagrang

The (pre)symplectic $\Psi$ and Poincar\'e-Cartan $\cal P$ submanifolds are
known as the phenomenological material relations
($p=mv,$ Kepler problem $f_A=-q^{-3}q_A,\,D=\varepsilon_0 E,\,B=\mu_0 H,$
London equation $J_\mu=A_\mu,\,$ etc) and are discussed in the last
sections.

Jacobi in 1838 proved that in mechanics the lagrangian and the
Hamilton-Jacobi submanifolds, $\Psi_2$ and $\Psi_4$, are the families of
solutions, $\phi\hr\Psi_4$ and $\phi\hr({\cal P}\circ\Psi_4)\hr\Psi_2.$ The
coordinate-free proof is in (Oziewicz and Gruhn 1983).
The extension of the Jacobi theorem beyond mechanics is not known.

The dimensions of submanifolds are given for the {\em phenomenological
field theory} (13) and follow from the considerations in the part II,
see formulas (17-18).

\begin{tabular}{r|ccccccc}
&E&$\hl$&$\Psi\hl{\cal P}$&$\hl$&$\Psi_4$&$\hl$&$\phi$\\
&\phantom{dim}$\downarrow$dim&&\phantom{dim}$\downarrow$dim&
&\phantom{dim}$\downarrow$dim&&\phantom{dim}$\downarrow$dim\\
mechanics&1+4n&&1+2n&&1+n&&1\\
strings&2+6n&&2+3n&&2+n&&2\\
electrostatics&3+8n&&3+4n&&3+n&&3\\
magnetostatics&3+12n&&3+6n&&3+3n&&3\\
Klein-Gordon fields&4+10n&&4+5n&&4+n&&4\\
electromagnetism&4+20n&&4+10n&&4+4n&&4\\
&\ss God's\\choice\kk&&\ss material\\relations\kk&&\ss
``quantum\\space"\kk&&\ss
time\\space\\space-time\kk\\
\hline\end{tabular}

\subsection*{Hamilton-Lagrange field theory}
\begin{vertical2}
The field theory $\{E,\eVer,\Omega\}$ is said to be $k$-{\em vertical} if
$0\neq d\Omega\in\Lambda_{(k)}$ and $d\Omega\not\in\Lambda_{(k-1)}$ or if
$d\Omega=0,$ $\Omega\in\Lambda_{(k)}$ and $\Omega\not\in\Lambda_{(k-1)}.$
\end{vertical2}
{\bf Comment.} For a cocycle $\Omega$ this definition was introduced by
Kondrack
(1978). The notion of the $k$-vertical field theory is essential for the
theory of the Poincar\'e-Cartan forms if $\Omega\not\in Z$ and for the
Hamilton-Jacobi theory if $\Omega\in Z.$

Note that $$\{\Lambda^{|\theta|+j}_{(k)}\}/(k-1)\neq 0\quad\mbox{iff}\quad
j\leq k\leq|\theta|+j.$$
In particular $\quad|\Omega|=1+|\theta|\quad\Ra\quad\Omega\cap\Lambda_{(0)}=0.$

In the section {\em calculus with the splitting} we are showing that (see the
definition 10 and corollary 3)
$$d\Lambda_{(k)}\subset\left\{\ba{clc}
\Lambda_{(k+1)}&\mbox{if Ver is integrable,}&d\theta\in\Lambda_{(1)}\\
\Lambda_{(k+2)}&\mbox{otherwise,}&d\theta\not\in\Lambda_{(1)}.
\ea\right..$$
Let the distribution Ver be integrable. In this case
$$\left\{\ba{l}d\Omega\in\Lambda_{(k)}\\d\Omega\not\in\Lambda_{(k-1)}\ea\right\}
\left\{\ba{l}\Omega\in\Lambda_{(k-1)}\oplus
Z\\\Omega\not\in\Lambda_{(k-2)}\oplu

The form $d\Omega\neq 0$ is $k$-vertical iff $\Omega$ can be decomposed (not
uni
$(k-1)$-vertical form and a cocycle.
Two fibrations of the de Rham complex $\Lambda$ of the differential forms are
involved in this decomposition: the first
over the factor ${\cal F}$-module $\Lambda/\Lambda_{(k-1)},$ the second over
the
$\Rint$-space $\Lambda/Z.$ The form $d\Omega$ is $k$-vertical if exist the
splittings
\begin{eqnarray*}
\Lambda/Z\quad&\stackrel{\mu}{\ra}&\Lambda,\\
\Lambda/\Lambda_{(k-1)}&\stackrel{\nu}{\ra}&\Lambda,
\end{eqnarray*} such that
$$\mu(\Omega/Z)\;\in\;\Lambda_{(k-1)},\qquad\nu(\Omega/(k-1))\;\in\;Z$$
\be\mbox{and}\qquad\Omega=\mu(\Omega/Z)+\nu(\Omega/(k-1))\;\in\;\Lambda_{(k-1)}\
Such splittings if exists are not unique, they are determined up to the
$(k-1)$-vertical cocycles $$\Lambda_{(k-1)}\cap Z
\;\stackrel{loc}{\simeq}\;d\Lambda_{(k-2)}.$$

Let the field theory $\{E,\Ver,\Omega\}$ be $k$-vertical.
Then the splitting $\mu$ determine $\nu$ and vice
versa. Locally
$$\nu(\Omega/(k-1))\;\stackrel{loc}{\simeq}\;d\omega,$$
\be\mbox{and}\quad\Omega=f+d\omega,\;\mbox{where}\;f\equiv\Omega-d\omega\,\in\,
\Lambda_{(k-1)}.\ee
The form $\omega$ is determined modulo $(k-2)$-vertical forms.
On Poincar\'e-Cartan submanifold, ${\cal P}\hr E,$ $\Omega$ and $f$ are exact,
\be dF\equiv{\cal P}^\ast f,\qquad d\alpha_F\equiv{\cal P}^{\ast}\Omega,\ee
$$\mbox{and}\qquad\alpha_F=F+{\cal P}^\ast\omega\quad\mbox{mod}\quad Z_{{\cal
P}
There exists the correlation between the decompositions (2-3) and
the Poin\-car\'e-Cartan submanifolds.

Depending on the choice of $f$ in (3), the potential $F$ could coincide (up to
t
sign) with the hamiltonian $H$ or with the lagrangian $L$ (see the next
sections
freedom in the decomposition (2-3) allows to see more possibilities.

If $f$ is $(k-1)$-vertical on $\{E,\Ver\},$ then $F$ (4) is
$(k-2)$-vertical on $\{{\cal P},{\cal P}^*\theta\},$
\be (\Lambda^{|\theta|}_{(k-2)}\oplus Z_{\cal P})\ni F\quad\longmapsto\quad
dF\i
The Poincar\'e-Cartan
equation (4), $dF={\cal P}^\ast f,$ allows to express ${\cal P}^\ast\omega$ in
terms of the partial derivatives of $F$ wrt the basis
of the ${\cal F}_{\cal P}$-module
$\Lambda^{|\theta|+1}_{(k-1)}\subset\Lambda_{\cal P}.$ Therefore
differential form $F$ determine the Poincar\'e-Cartan form,
$$\Lambda^{|\theta|}_{(k-2)}/Z_{\cal P}\ni
F\quad\longmapsto\quad\alpha_F\in\Lam
P}^{|\theta|}\quad\mbox{mod}\;Z_{\cal P}.$$
This motivate the definition
\begin{HL} Let the distribution {\eVer} be integrable and $\Omega\not\in Z.$
\bd\item{(i)} Let $2\leq k\leq |d\Omega|$ and let the field theory
$\{E,{\eVer},
$d\Omega\in\Lambda_{(k)}$ and
$d\Omega\not\in\Omega_{(k-1)}$. Then the ${\cal F}_{\cal P}$-module
$\Lambda^{|\theta|}_{(k-2)}$ is said to be the module {\em generating} the
Poincar\'e-Cartan forms.
\item{(ii)} The field theory $\{E,\theta,\Omega\}$ is said to be the
{\em Hamilton-Lagrange} field theory, abbreviated by {\em HL}, if the ${\cal
F}_
P}$-dimension of the generating module of the Poincar\'e-Cartan forms is 1.
\ed\end{HL}
For HL field theory the Poincar\'e-Cartan form is determined by {\em one}
(pseudo)\-sca\-lar function, (lagrangian,
hamiltonian,$\dots$).
\begin{HL-type} The field theory $\{E,{\eVer},\Omega\not\in Z\}$ is {\em HL}
iff
$d\Omega$ is bi-vertical, $d\Omega\in\Lambda_{(2)}.$
\end{HL-type}
\noindent{\bf Proof.}
$$\dim
\{\Lambda^{|\theta|}_{(k-2)}\}=\sum_{i=0}^{min(k-2,|\theta|)}\choose{\dim\,\Ver}
Therefore
$$\dim\{\Lambda^{|\theta|}_{(k-2)}\}=1\quad\mbox{iff}\quad\{E,{\Ver},\Omega\not\
Z\}\;\mbox{is
bi-vertical}.$$
{\bf Comment.} In HL field theory the (local) hamiltonian
and lagrangian $F$ (4) are the {\em vertical} differential forms on $\cal P.$
For not HL-type theories the analogous hamiltonian or lagrangian differential
forms are no more vertical and therefore can not be expressed by means of one
pseudoscalar function. Analogous considerations are valid for the
Hamilton-Jacobi theory if $\Omega\in Z.$

\vspace{0.1cm}\noindent{\bf Partial derivatives of vertical forms.}
Note that $$\dim\{\Lambda^{|\theta|+1}_{(1)}\}=\dim\,\Ver.$$
We will suppose that the modul $\Lambda^{|\theta|+1}_{(1)},$ on $E$ as well as
o
submanifold ${\cal P},$ is generated by the differentials of the homogeneous
vertical forms (of different degrees),
$$\Lambda^{|\theta|+1}_{(1)}\equiv\mbox{gen}\{dw^A,\,w^A\in\Lambda_{(0)}\}.$$
This means that $\forall\:\alpha\in\Lambda^{|\theta|+1}_{(1)}$ has the unique
de
$$\alpha=dw^A\wedge\alpha_A,\qquad w^A,\alpha_A\in\Lambda_{(0)}.$$
In particular the generating set $\{dw^A\}$ determine the partial derivatives
of
the highest degree vertical forms
$$\Lambda^{|\theta|}_{(0)}\ni F\longmapsto dF=dw^A\wedge\frac{\partial
F}{\partial w^A}\in\Lambda^{|\theta|+1}_{(1)}.$$
Example
$$dL\equiv dq^A\wedge\frac{\partial L}{\partial q^A}+dv^A\wedge\frac{\partial
L}{\partial v^A}.$$

\section*{II. Phenomenology}
\subsection*{Phenomenological field theory}
Let $\{q^A,v^A,p_A,f_A;\,A\in I\subset\Nint\}$ be a collection of the
vertical differential forms on $\{E,{\Ver}\}.$ Newton's and Maxwell's
phenomenological equations as well as of electrostatics and magnetostatics have
the form
\begin{eqnarray}
\phi^*(dq^A-v^A)&=&0\nonumber\\
\phi^*(dp_A\,-f_A)&=&0.
\end{eqnarray}
The differential forms $\{q^A,v^A,p_A,f_A\}$ in (6) are {\em
independent}, as they are determined by independent experiments.
The phenomenological material relations among these fields are the consequence
of the further independent measurements. This was stressed by Newton (1686) and
Maxwell. After Lagrange it became customary to present the Newton's equations
(as well as of electrodynamics, $\Box A=j$) as the second order from the
begining, contrary to the original Newton's presentation.
Also Piron stressed (e.g. in Piron's {\em Lectures on electrodynamics}, 1989)
that the equations (6) should not presuppose the material relations.

The phenomenological material relations are the equations for the
(pre)\-sym\-plec\-tic submanifold of the field
theory $\{E,{\Ver},\Omega\}$ (definition 4 (i)), and are discussed in the next
sections.

The aim is to determine the most general {\em regular} field theory
$\{E,{\Ver},\Omega\}$ which field equations (1) coincide with the experimental
one (6). The diffe\-rent field theories with the same set of the {\em first}
ord
equations (6), will leads to the different (pre)symplectic submanifolds and
therefore to the different {\em second} order equations
considered in the part III.
Denote\begin{eqnarray}
\vartheta^A&\equiv&dq^A-v^A\nonumber\\
\omega_A&\equiv&dp_A-f_A.\end{eqnarray}
The solutions $\phi$ (6) anihilate the ideal generated by
$\{\vartheta^A,\omega_A\},$ therefore for the regular field theory we need
the equality of the ideals,
\be \mbox{gen}\{i_{\mbox{{\scriptsize
Ver}}}\Omega\}\;=\;\mbox{gen}\{\vartheta^A
If the distribution Ver is integrable then $\{\vartheta^A,\omega_A\}$ are
$1$-vertical.
The most general form $\Omega$ compatible with (8) needs to be $2$-vertical of
t
\be
\Omega\equiv\sum_{A,B}(K^A_B\wedge\omega_A\wedge\vartheta^B+
\Gamma_{AB}\wedge\vartheta^A\wedge\vartheta^B+\chi^{AB}\wedge\omega_A\wedge\omeg
where $\{K^A_B,\Gamma_{AB},\chi^{AB}\}$ is the collection of a vertical
differential forms such that
$$\Gamma_{AB}=(-1)^{|\vartheta^A||\vartheta^B|}\Gamma_{BA},\qquad\chi^{AB}=
(-1)^{|\omega_A||\omega_B|}\chi^{BA}.$$
Define the {\em dimension} of the form as the dimension of the factor module,
\be \dim\alpha\equiv\dim\,\Ver-\dim\{(\mbox{Dist}\,\alpha)\cap\Ver\}.\ee
One of the necessary condition for the implication (9) $\Rightarrow$ (6) is
\be \dim\Ver=\sum_A(\dim\,dq^A+\dim\,dv^A+\dim\,dp_A+\dim\,df_A).\ee
\begin{pft}
The field theory $\{E,{\eVer},\Omega\}$ with $\Omega$ of the form (9) is said
to be the {\em phenomenological field theory}.
\end{pft}
Because $\Omega$ is $\Nint$-homogeneous
then
\begin{eqnarray}
|v^A|&=&1+|q^A|\nonumber\\
|f_A|&=&1+|p_A|\nonumber\\
|q^B|+|f_A|+|K^A_B|&=&|\theta|\nonumber\\
|v^B|+|p_A|+|K^A_B|&=&|\theta|
\end{eqnarray}
In mechanics $|K|=|\Gamma|=|\chi|=0.$ The conditions
$$\forall\;A\quad|\omega_A|\geq|\vartheta^A|\quad\mbox{and}\quad|K|=|\chi|=0,$$
determine the unique grades for mechanics and string theory
and $n$ possibilities for $|\theta|=2n-1$ and $2n.$
In this case $\chi$ can contribute
in mechanics and magnetostatics only.
$$\ba[t]{ccccccl}
|\theta|&|q|&|v|&|p|&|f|&|\Gamma|&\\\hline
1&0&1&0&1&0&\mbox{mechanics}\\
2&0&1&1&2&1&\mbox{strings}\\
3&0&1&2&3&2&\mbox{electrostatics}\\
&1&2&1&2&0&\mbox{magnetostatics}\\
4&0&1&3&4&3&\mbox{Klein-Gordon scalar fields}\\
&1&2&2&3&1&\mbox{electromagnetic field}\\
\hline\ea$$
If $\{K,\Gamma,\chi\}$ are the vertical cocycles then the field theory (13) is
HL, $d\Omega$ is bivertical, and
$$\Omega=\{d(K^B_A\wedge p_B+\psi\Gamma_{AB}\wedge q^B)-(K^B_A\wedge
f_B+\Gamma_
v^B)\}\wedge\vartheta^A+\omega^A\wedge\omega_A.$$
Effectively the ``momenta-induction" and ``force-current" are rotated and
translated by ``connection $\Gamma$", a natural description of the
velocity-dependent forces,
\begin{eqnarray*}
p_A&\longmapsto&K^B_A\wedge p_B+\psi\Gamma_{AB}\wedge q^B,\\
f_A&\longmapsto&K^B_A\wedge f_B\,+\;\;\Gamma_{BA}\wedge v^B.
\end{eqnarray*}
The phenomenological {\em symplectic mechanics} (9) without of the $\chi$-terms
has been considered by Jadczyk and Modugno (1992).

Consider HL field theory
\be\Omega\equiv\sum\omega_A\wedge\vartheta^A\quad\in\;\Lambda_{(1)}\oplus Z
\quad\subset\Lambda_{(2)}.\ee

The $1$-vertical differential forms
$\{h,l,s,t\}$ are defined by the decompositions (see formulas (2-3)),

\begin{eqnarray}
\Omega&=&-h+d(p_A\wedge dq^A)\nonumber\\
&=&+l+d\{p_A\wedge(dq^A-v^A)\}\nonumber\\
&=&+s+d\{q^A\wedge\psi^{|\theta|}(dp_A-f_A)\}\nonumber\\
&=&+t+d\{q^A\wedge\psi^{|\theta|}(dp_A-f_A)-p_A\wedge v^A\}.
\end{eqnarray}
Explicitely
\begin{eqnarray}
h&\equiv&dp_A\wedge v^A\qquad\:+dq^A\wedge\psi^{|\theta|}f_A,\nonumber\\
l&\equiv&dv^A\wedge
\psi^{1+|\theta|}p_A-dq^A\wedge\psi^{|\theta|}f_A,\nonumber\\
s&\equiv&dp_A\wedge v^A\qquad\;-df_A\wedge(-)^{|\theta|}\psi q^A,\nonumber\\
t&\equiv&dv^A\wedge \psi^{1+|\theta|}p_A+df_A\wedge(-)^{|\theta|}\psi q^A.
\end{eqnarray}
{\bf Remark.}
\begin{eqnarray}
l+h\equiv t+s&\equiv&d(\,p_A\wedge v^A),\nonumber\\
h-s\equiv t-l&\equiv&d(\psi f_A\wedge q^A).
\end{eqnarray}

If $E$ is fibered and the differential forms $\{q^A,v^A,p_A,f_A\}$
are the Louville's forms, considered in the next section, then the
differential form $\Omega$ (13) is regular and imply the
field equations (6).

\subsection*{Louville's differential forms}
Let $M$ be a manifold and for $p\in M,$ $T_p^*M$ be the vector $I\!\!R$-space
of exterior forms at $p.$ Let $T^kM\stackrel{\pi}{\ra}M,$ be the bundle of
exterior $k$-forms, $(T^kM)_p\equiv(T^*_pM)^{\wedge k},$ with $(T^*_pM)^{\wedge
0}\equiv I\!\!R.$

The differential form $\alpha\in\Lambda^k_M$ determine the section
$\alpha_s\in\Gamma(M,T^kM),$
$\Lambda_{T^kM}\ni\lambda\ra\alpha_s^*\lambda\in\Lambda_M.$
\begin{Louvill} The differential $k$-form $\lambda\in\Lambda_{T^kM}$ is  said
to
be the Louville's form if
$$\alpha_s^*\lambda=\alpha\quad\mbox{for every}\quad\alpha\in\Lambda^k_M.$$
\end{Louvill}
There exists the unique Louville's differential $k$-form and has the local form
$$\lambda=\frac{1}{k!}\sum\lambda_{\mu_1\dots\mu_k}\pi^*(dt^{\mu_1}\wedge\dots\w
dt^{\mu_k}).$$
The Louville's differential forms of arbitrary degree has been introduced by
Tulczyjew in 1979. The Louville's forms are vertical wrt
$\theta\equiv\pi^*\vol_

Let the manifold $E$ be the bundle, $E\stackrel{\pi}{\ra}M,$ of the exterior
forms of different degrees on the manifold $M,$
\be E\equiv
\bigoplus_A\left\{(T^{|q^A|}M)\oplus(T^{|v^A|}M)\oplus(T^{|p_A|}M)\oplus(T^{|f_A
Let $\{q^A,v^A,p_A,f_A\}$ be a collection of the Louville's forms on $E.$

If $\lambda$ is a Louville's differential form on $E$ then according to the
definition (14-15),
\be \dim\,(d\lambda)=\choose{\dim\,M}{|\lambda|}.\ee
The field theory $\{E\,(17),\pi^*\vol_M, \Omega\,(13)\}$ is regular and imply
th
field equations (6).
{}From formula (11) we get the dimensions, dim$E$, listed in the Table after
the
definition 4.
The Louville's forms $\{K,\Gamma,\chi\},$ for simplicity, are not included in
$E$ (17). The Louville's forms $\{K,\Gamma,\chi\}$ in (9) contribute to dim$E.$
In mechanics with (9), $|\theta|=1,$ dim$E=1+4n+2n^2-n.$

\subsection*{Poincar\'e-Cartan submanifolds}
Consideration of this section are the same for the case of presymplectic and
Poincar\'e-Cartan submanifolds. To be specific we will consider only
Poincar\'e-Cartan submanifolds, definition 4 (ii), and only for the case when
$E$ is a bundle of exterior forms (17) with the conditions (12).

Let the bundle $E$ be splitted with the fiber-preserving projectors
$\pi_{P/C},$
$$\ba{ccccc}
P\phantom{\pi}&\stackrel{\pi_P}{\la}&E=P\oplus
C&\stackrel{\pi_C}{\ra}&C\phantom{\pi}\\
\downarrow\pi&&\phantom{E=}\downarrow\pi&&\downarrow\pi\\
M\phantom{\pi}&=&\phantom{E=}M\phantom{\pi}&=&M\phantom{\pi}\ea$$
Let the subbundles $P$ and $C$ be of equal dimensions, dim$\,$Ver (14) on $E$
is
The Poincar\'e-Cartan submanifold ${\cal P}\hr E$ is a subbundle of $E$
with dim${\cal P}=\dim\,P=\dim\,C.$ Let $\pi_P|{\cal P}$ be the
fiber-preserving isomorphism. Then ${\cal P}$ will be identified with the
injection ${\cal P}:P\hr{\cal P}\subset E,$ $\pi_P\circ{\cal P}=\id_P,$ and
$\varphi\equiv\pi_C\circ{\cal P}:P\ra C,$ is a fiber-preserving bundle map.
If $\theta\equiv\pi^\ast\vol_M$ then ${\cal
P}^\ast\theta\equiv\theta\;\in\;\Lam

Consider the splittings of the bundle $E$ for which $\Omega$ have the
form (compare with the decomposition (2-3))
\be\Omega=\pm\sum d\pi_P^*\alpha\wedge\pi_C^*\beta+d\omega.\ee
On Poincar\'e-Cartan submanifold the form $\sum
d\pi_P^*\alpha\wedge\pi_C^*\beta$ is exact,
$${\cal P}^*(\sum
d\pi_P^*\alpha\wedge\pi_C^*\beta)=\sum d\alpha\wedge\varphi^*\beta\equiv
dF\;\in\;\Lambda_P.$$
Therefore
$$\varphi^*\beta\equiv\frac{\partial F}{\partial \alpha}.$$

The differential forms $\{\Omega,h,l,s,t\}$ (12-15) are exact on
Poincar\'e-Cart
submanifold ${\cal P}\hr\Psi\hr E.$ In particular the hamiltonian $H$ and the
lagrangian $L$ are the differential forms on {\em different} Poincar\'e-Cartan
subbundles and are defined as the
potentials,
\be
dH\equiv{\cal P}^*_hh,\quad
dL\equiv{\cal P}^*_ll,\quad
dS\equiv{\cal P}^*_ss,\quad
dT\equiv{\cal P}^*_tt.\ee
The compositions, say ${\cal L}\equiv{\cal P}^{-1}_h\circ{\cal P}_l$ etc, are
sa
to be the {Legendre's transforms}. With the help of the identities (16) the
Legendre's transforms allow to calculate, for example, the lagrangian $L$ for
th
given hamiltonian $H,$
\begin{eqnarray}
dL={\cal P}^*_ll&=&({\cal P}_l^*\circ{\cal P}_h^{-1*}\circ{\cal
P}^*_h)\{d(p_A\wedge v^A)-h\}\nonumber\\&=&{\cal L}^*\{d{\cal P}^*_h(p_A\wedge
v
\end{eqnarray}
Therefore, modulo cocycles
\be L=({\cal L}^*\circ{\cal P}_h^*)(p_A\wedge v^A)-{\cal
L}^*H\quad\mbox{mod}\;Z

The Poincar\'e-Cartan forms $\alpha,\,d\alpha\equiv{\cal
P}^*\Omega,$ can be expressed in terms of $L,H,S,$ and $T$ if we identify the
decompositions (14-15) with (19-20).

\vspace{0.5cm}
$$\varphi_h^\ast v^A\equiv\frac{\partial H}{\partial p_A},
\qquad\varphi_h^\ast f_A\equiv
\psi^{|\theta|}\frac{\partial H}{\partial q^A},$$
$$\alpha_H\equiv-H+p_A\wedge dq^A\quad\mbox{mod}\,Z_{\cal P},$$
\be d\alpha_H=\left(dp_A-\psi^{|\theta|}\frac{\partial H}{\partial q^A}\right)
\wedge\left(dq^A-\frac{\partial H}{\partial p_A}\right).\ee

\vspace{0.5cm}$$
\varphi_l^\ast p_A\equiv\psi^{(1+|\theta|)}\frac{\partial L}{\partial
v^A},\qquad\varphi_l^\ast f_A\equiv-\psi^{|\theta|}\frac{\partial L}{\partial
q^
$$\alpha_L\equiv L+\frac{\partial L}{\partial
v^A}\wedge\psi^{(1+|\theta|)}(dq^A-v^A)\quad\mbox{mod}\,Z_{\cal P},$$
\be d\alpha_L=\left\{d\frac{\partial
L}{\partial v^A}-(-)^{|\theta|}\psi\frac{\partial L}{\partial
q^A}\right\}\wedge\psi^{(1+|\theta|)}(dq^A-v^A).\ee

\vspace{0.5cm}$$
\varphi_s^\ast v^A\equiv\frac{\partial S}{\partial p_A},
\qquad\varphi_s^\ast q^A\equiv
(-)^{1+|\theta|}\psi\frac{\partial S}{\partial f_A},$$
$$\alpha_S\equiv S+\frac{\partial S}{\partial
f_A}\wedge\psi^{(1+|\theta|)}(dp_A-f_A)\quad\mbox{mod}\,Z_{\cal P},$$
\be d\alpha_S=-\left\{d\frac{\partial S}{\partial
f_A}-(-)^{|\theta|}\psi\frac{\partial S}{\partial
p_A}\right\}\wedge\psi^{(1+|\theta|)}(dp_A-f_A).\ee

\vspace{0.5cm}$$
\varphi_t^\ast p_A\equiv\psi^{(1+|\theta|)}\frac{\partial T}{\partial
v^A},\qquad\varphi_t^\ast q^A\equiv(-)^{|\theta|}\psi\frac{\partial T}{\partial
$$\alpha_T\equiv T+\left\{d\psi^{1+|\theta|}\frac{\partial T}{\partial
v^A}-f_A\
\wedge\frac{\partial T}{\partial f_A}-v^A\wedge\frac{\partial T}{\partial v^A}
\quad\mbox{mod}\,Z_{\cal P},$$
\be d\alpha_T=\left\{(-)^{1+|\theta|}d\frac{\partial T}{\partial f_A}-\psi
v^A\r
\wedge\left\{d\frac{\partial T}{\partial
v^A}-(-\psi)^{1+|\theta|}f_A\right\}.\e
The $(f,v)$-subbundle (26) in electromagnetism was considered by Thirring
(1979,
109) and therefore one is tempted to call the differential form $T$ as the
{\em thirringian}. See also the last two sections in the part III.

\section*{III. Field equations}
\subsection*{The Hodge's map}
We give slight generalization of the
Hodge's map for the manifold with the distribution Ver$\,\subset W.$
When $\theta=\vol\;(\Leftrightarrow \mbox{Dist}\theta\equiv\Ver=0)$ then this
ge\-ne\-ralization
collapse to the usual one. The formulas in the lemma 4 do not suppose that the
scalar product $g$ is symmetric.

By definition the form $\theta$ is decomposable therefore
$$\cVer=\{\alpha\in\Lambda^1_E,\;\alpha\wedge\theta=0\}.$$
\begin{annul}
If $\alpha\in\cVer^\wedge$ then for each multivector $X\in W^\wedge$
$$\alpha\wedge i_X\theta=(-)^{|\alpha|(1+|X|)}i_{i_\alpha X}\theta.$$
\end{annul}
If $\alpha\in\cVer^\wedge$ and $|\alpha|=|X|$ then
$$\alpha\wedge i_X\theta=(\alpha X)\theta.$$

Consider the factor module $W/\Ver$ as the dual to $\cal V$er and let
$Z\in(W/\Ver)^{\wedge |\theta|}$ be such that $\theta Z=1.$

Let $g,\,k$ and their pull-backs (a transpositions) $g^*,\,k^*$ be ${\cal
F}$-linear maps
\begin{eqnarray*}
g,\,g^*:\:\;\quad\,\cVer\theta\,&\ra& W/\Ver,\\
k,\,k^*:\quad W/\Ver&\ra&\cVer\,\theta.
\end{eqnarray*}
For $\alpha$ and
$\beta$ in $\cVer,$ $g(\alpha\otimes\beta)=g^*(\beta\otimes\alpha),$ $g$ and
$k$
need not to be symmetric (as in Oziewicz 1986).
The extension of maps (28) to Grassmann algebra maps
$(\cVer)^\wedge\ra(W/\Ver)^\wedge$ is denoted by the same letters.
\begin{Hodge}
The ${\cal F}$-linear maps $\ast\equiv \ast_{(g,\theta)}$ and
$\star\equiv\star_{(k,Z)},$
\begin{eqnarray*}
(\cVer)^\wedge\ni\alpha&\longmapsto& \ast\alpha\equiv
i_{g\alpha}\theta\;\,\in\cVer^\wedge,\\
(\cVer)^\wedge\ni\alpha&\longmapsto& \star\alpha\equiv
ki_\alpha Z\in\cVer^\wedge,
\end{eqnarray*}
are said to be the Hodge's maps with $\theta\equiv\ast 1$ and
$\,\star\theta\equ
\end{Hodge}
\begin{wHodge}.\samepage
\begin{description}
\item{(i)}\hspace{4.0cm}  $\ast_g\circ e_\alpha=i_{g\alpha}\circ
\ast_g\circ\psi
\item{(ii)}\hspace{3.5cm} $\;\ast_{g^*}\circ
i_{g\alpha}=e_{\psi\alpha}\circ\ast
\item{(iii)}\hspace{4.5cm} $\psi\circ \ast=(-)^{|\theta|}\ast\circ\psi.$
\item{(iv)}\hspace{3.5 cm} $
\ast_{(g,\theta)}\circ\ast_{(g^*,\theta)}=g(\theta\otimes\theta)\circ\psi^{1+|\t
}.$
\item{(v)} If $\alpha$ and
$\beta\in(\cVer\,\theta)^\wedge$ are of equal grades, then
$$\alpha\wedge\ast\beta=g(\beta\otimes\alpha)\theta=g^*(\alpha\otimes\beta)\thet
=\beta\wedge\ast_{g^*}\alpha,$$
$$g(\ast\alpha\otimes\ast\beta)=g^*(\theta\otimes\theta)\cdot
g(\alpha\otimes\beta).$$
\end{description}
\end{wHodge}
The function $g(\theta\otimes\theta)\in{\cal F}_E$ is said to be the
$\theta$-determinant of $g,$ $${{\det}_{\theta}}g\equiv
g(\theta\otimes\theta)\e
Let $\theta_g\equiv|g(\theta\otimes\theta)|^{-1/2}\theta,$ then
$g(\theta_g\otimes\theta_g)=\mbox{sign}\det_{\theta}g.$

\subsection*{Calculus with the splitting}
Let $\forall\,X\in W,$
$\nabla_X$ be a zero-grade derivation of the tensor algebra such that
$\nabla_X|{\cal F}\equiv X.$ The derivation $\nabla_X$ factors to the
derivation of exterior algebra, $\nabla_X\in\der\Lambda.$ For
form $\alpha,$ the composition $e_\alpha\circ\nabla_X$ is a (skew) derivation
of $\Lambda,$ $|e_\alpha\circ\nabla_X|=|\alpha|.$
Let $\Lambda^1\equiv\mbox{span}\{\varepsilon^a\}$ and
W$\equiv\mbox{span}\{X_a\}
where $\varepsilon^a X_b\equiv\delta^a_b.$
Let $\nabla_a\equiv\nabla_{X_a},$ then
$\nabla\equiv\varepsilon^a\wedge\nabla_a$ is a skew derivation, $|\nabla|=+1$
and $\nabla|{\cal F}\equiv d.$ The difference $\nabla-d$ is said to
be the torsion of $\nabla,$
$$\nabla=d+T\quad\in\quad\mbox{skew}\,\der\Lambda.$$
The splitting of the Definition 3 (iii) determine the bigradation
($\Nint\times\Nint$-grading) of $\Lambda=\oplus\Lambda^{p,q},$ with
the projectors $\pi^{p,q},$ $\Lambda^{p,q}\equiv\pi^{p,q}(\Lambda^{p+q}).$
Every
derivation of grade $+1,$ is determined by values on generators $\{{\cal
F},d{\cal F}\},$ and therefore is decomposed as the sum of four bigraded
derivations.
\begin{bigraded} Let {\eVer}$\,\equiv\pi^{1,0}\Lambda^1,$
{\eHor}$\,\equiv\pi^{0,1}\Lambda^1$ and {\em bigrade} $\equiv\,|\cdot|.$
$$\ba{lll}
d_V|{\cal F}\equiv \pi^{1,0}\circ d,&
d_V|d{\cal F}\equiv\pi^{2,0}\circ d\circ\pi^{1,0}+\pi^{1,1}\circ
d\circ\pi^{0,1},&|d_V|\phantom{_n}=(+1,\phantom{+}0),\\
d_{nV}|{\cal F}\equiv 0,&d_{nV}|d{\cal F}\equiv\pi^{0,2}\circ
d\circ\pi^{1,0},&|d_{nV}|=(-1,+2),\\
d_H|{\cal F}\equiv\pi^{0,1}\circ d,&
d_H|d{\cal F}\equiv\pi^{0,2}\circ d\circ\pi^{0,1}+\pi^{1,1}\circ
d\circ\pi^{1,0},&|d_H|\phantom{_n}=(\phantom{-}0,+1),\\
d_{nH}|{\cal F}\equiv 0,&d_{nH}|d{\cal F}\equiv\pi^{2,0}\circ
d\circ\pi^{0,1},&|d_{nH}|=(+2,-1).\\
\ea$$
\end{bigraded}
Note that $d_V:{\cal F}\rightarrow \cVer$ and $d_H:{\cal F}\rightarrow {\cal
H}or.$ The distribution Ver is integrable iff $d_{nV}=0$ and the distribution
Hor is integrable iff $d_{nH}=0.$ The tensors $d_{nV}$ and $d_{nH}$
measure the {\em non}involutiveness of these distributions.
\begin{seven}
\be d=d_{nH}+d_V+d_H+d_{nV},\ee
and $d^2=0$ is equivalent to {\em seven} conditions:
$$\begin{array}{rl}
d_{nV}\circ d_{nV}=0,&\quad\ebigrade\,=\,(-2,+4)\\
d_H\circ d_{nV}+d_{nV}\circ d_H=0,&\quad\ebigrade\,=\,(-1,+3)\\
d_V\circ d_{nV}+d_{nV}\circ d_V+d_H^2=0,&\quad\ebigrade\,=\,(\phantom{-}0,+2)\\
d_{nV}\circ d_{nH}+d_{nH}\circ d_{nV}+d_V\circ d_H+d_H\circ
d_V=0,&\quad\ebigrade\,=\,(+1,+1)\\
d_H\circ d_{nH}+d_{nH}\circ d_H+d_V^2=0,&\quad\ebigrade\,=\,(+2,\phantom{+}0)\\
d_V\circ d_{nH}+d_{nH}\circ d_V=0,&\quad\ebigrade\,=\,(+3,-1)\\
d_{nH}\circ d_{nH}=0,&\quad\ebigrade\,=\,(+4,-2).
\end{array}$$
\end{seven}
If the differential form $\alpha$ is vertical then $d_V\alpha$ is vertical,
$d_H\alpha$ is 1-vertical and $d_{nV}\alpha$ is 2-vertical. From the
definition 1 it follows for the phenomenological field theory
$\{E,\Ver,\Omega\}$ (13) that
$$\Omega\quad\mbox{is}\quad\left\{\begin{array}{ll}
2-vertical\,&\mbox{if Ver is integrable,}\\
4-vertical&\,\mbox{otherwise.}\end{array}\right.$$

\subsection*{Dirac operator and codifferential}
Let $g:\cVer\rightarrow$Hor, and
let $\gamma:\cVer^\wedge\ra\lin(\cVer^\wedge)$ be the left Clifford
multiplication, for $\varepsilon\in\cVer$ and $\alpha\in(\cVer)^\wedge,$
 \be \gamma_\varepsilon\alpha\equiv\varepsilon\wedge\alpha
+i_{g\varepsilon}\alpha,\ee
(e.g. Oziewicz 1986). Denote $\gamma^\mu\equiv\gamma(\varepsilon^\mu).$
\begin{Dirac}
The operator $D\equiv\gamma^\mu\circ\nabla_\mu:\;\Lambda\rightarrow\Lambda$
is said to be the splitting-dependent Dirac operator.
\end{Dirac}
When Ver$\,=0$ then $D$ collapse to the usual Dirac operator.

Let
$\cVer\equiv\mbox{span}\{\varepsilon^\mu\}$ and
Hor$\equiv\mbox{span}\{h_\mu\},$
where $\varepsilon^\mu h_\nu\equiv\delta^\mu_\nu.$ Then
$$
\heartsuit
\equiv\varepsilon^\mu\wedge\nabla_\mu\quad\in\quad\mbox{skew}\,\der\L
$$ \be\heartsuit=d_V+T_\heartsuit.\ee
Because of (27) we have $D=\heartsuit+\delta$ where $\delta$ is a
(splitting-dependent) codifferential,
$$\delta\equiv i_{g\varepsilon^\mu}\circ\nabla_\mu,\qquad|\delta|=-1.$$
Using Lemma 4 [(i) and (ii)] we calculate
\begin{eqnarray*}
\delta\circ \ast_g\quad\,&=&+\ast_g\circ
\heartsuit\circ\psi+i_{g\varepsilon^\mu}\circ(\nabla_\mu\ast_g),\\
\ast_{g^*}\circ \delta\circ\psi&=&-\:\heartsuit\circ
\ast_{g^*}\quad+e_{\varepsi
\ast_{g^*}).
\end{eqnarray*}
Therefore
$$\delta=(\mbox{sign}\,{\det}_{\theta} g)(\ast_{(g,\theta_g)}\circ
\heartsuit\circ \ast_{(g^*,\theta_g)}\circ
(-\psi)^{|\theta|}+``\nabla\ast"),$$
where $$``\nabla\ast"\equiv i_{g\varepsilon^\mu}\circ(\nabla_\mu
\ast_{(g,\theta_g)})\circ
\ast_{(g^*,\theta_g)}\circ \psi^{1+|\theta|}.$$
If the Hodge's map is parallel, $\nabla\ast=0,$ and $T_\heartsuit=0$ (28),
then $\heartsuit=d_V$ and the codifferential has the form
\be \delta=({{\mbox{sign}\det}_{\theta}}g)\ast_{(g,\theta_g)}\circ d_V\circ
\ast_{(g^*,\theta_g)}\circ (-\psi)^{|\theta|}.\ee
$$\delta^2=-\frac{\det g^*}{|\det g|}\ast_g\circ
d_V^2\circ\ast_g\circ\psi^{1+|\theta|}.$$
If Hor is integrable then $d_V^2=\delta^2=0.$ The square of the Dirac operator
${\triangle}_g\equiv
(d_V+\delta)^2=\delta\circ d_V+d_V\circ\delta$ is said to be the
(splitting-dependent) Laplace-Beltrami-Hodge operator, which commutes with the
Hodge's map if $g$ is symmetric,
$${\triangle}_g\circ\ast=\ast\circ{\triangle}_{g^*}.$$

\subsection*{Legendre's transforms}
Let $\{q^A,q_A,\ldots\}$ be the collection of the vertical forms,
$|q_A|=|q^A|,$
and let
\be q^2\equiv |g\sum q_A\otimes q^A|,\quad l\equiv l(q^2).\ee
With the help of formula (v) in the lemma 4 we calculate
\be d(l\sum q^A\wedge\ast q_A)=(q^2l)'d\sum(q^A\wedge\ast q_A)-q^2l'\sum
g(q_A\otimes q^A)d\theta.\ee
If $f,g\in{\cal F}$ then
$$d(f\wedge\ast g)=df\wedge\ast d+dg\wedge\ast f+fgd\theta.$$
This is no more valid for arbitrary grade. The formula (v) of lemma 4 suggest
th
simplifying assumption
\be \phantom{assumption}
d(\alpha\wedge\ast_g\beta)\simeq
d\alpha\wedge\ast_g\beta+d\beta\wedge\ast_{g^*}
This could be the case if $\theta$ is a cocycle and $\alpha$ and $\beta$
are the vertical differential forms on which $d=d_V.$
Let $\theta$ be a cocycle. With the help of (31-32) we get
\be d(l\sum q^A\wedge\ast q_A)\simeq 2(q^2l)'\sum dq^A\wedge\ast q_A.\ee
Every vertical form of highest grade can be expressed in terms of the
independent vertical forms. In particular the hamiltonian,
$H\in\Lambda^{|\theta|}_{(0)},$ can be expressed in terms of $\{p_A,q^A\}.$
Consider the example
\be H\equiv \frac{1}{2}k(p^2)\sum p_A\wedge\ast p^A+\frac{1}{2}l(q^2)\sum
q^A\wedge\ast q_A.\ee
More general expressions are possible. Looking at
the table after formulas (12) we see, for example, that magnetostatics offers
mo
possibilities than electrostatics for which the above expression (34) for the
hamiltonian is unique up to the scalar factors.

{}From (33-34) it follows $$\frac{\partial H}{\partial p_A}\simeq(p^2k)'\ast
p^A\quad\mbox{and}\quad\frac{\partial H}{\partial q^A}\simeq (q^2l)'\ast q_A.$$
With the help of the Legendre's transforms among different Poincar\'e-Cartan
subbundles of $E$ and with the help of the identities (16) we can calculate
the vertical differential forms $L,$ $S$ and $T$ (20) on other
Poincar\'e-Cartan
subbundles. For example
$$L={\cal L}^*(p_A\wedge\frac{\partial H}{\partial p_A}-H),\quad\mbox{where}\;
{\cal L}\equiv{\cal P}^{-1}_h\circ{\cal P}_l.$$
We will assume that the Legendre's transforms commute with the Hodge's map
$${\cal L}^*\circ\ast\simeq\ast\circ{\cal L}^*.$$
Then it is easy to calculate the Legendre's transform for monomials,
\be k(p^2)\equiv\frac{p^{2k}}{m}\quad\mbox{and}\quad l(q^2)\equiv
cq^{2l},\quad\mbox{where}\quad m,c\in\Rint.\ee
Denote
\begin{eqnarray*}
m_k&\equiv&
(\mbox{sign}\det\,g)|*\theta|^{-(k+1)/(2k+1)}\frac{1+2k}{1+k}(\frac{m}{1+k})^{1/
c_l&\equiv&(\mbox{sign}\det\,g)|*\theta|^{(l+1)/(2l+1)}\frac{1+l}{1+2l}[(1+l)c]^
\end{eqnarray*}
For the given hamiltonian (34-36) we get
\begin{eqnarray*}
H&\equiv&\frac{p^{2k}}{2m}\sum p_A\wedge\ast
p^A\hspace{2.0cm}+\frac{cq^{2l}}{2}
q^A\wedge\ast q_A,\\
L&=&\frac{m_k}{2}v^{-(2k)/(2k+1)}\sum v^A\wedge\ast v_A-\frac{cq^{2l}}{2}\sum
q^A\wedge\ast q_A,\\
S&=&\frac{p^{2k}}{2m}\sum p_A\wedge\ast
p^A\hspace{2.0cm}-\frac{1}{2c_l}f^{-(2l)
f_A\wedge\ast f^A,\\
T&=&\frac{m_k}{2}v^{-(2k)/(2k+1)}\sum v^A\wedge\ast
v_A-\frac{1}{2c_l}f^{-(2l)/(2l+1)}\sum f_A\wedge\ast f^A.
\end{eqnarray*}
Assuming that $m_k$ and $c_l$ are constant we have
$$
\begin{array}{ll}
\frac{\partial H}{\partial p_A}\simeq\frac{k+1}{m}p^{2k}\ast p^A,&
\frac{\partial H}{\partial q^A}\simeq (l+1)cq^{2l}\ast q_A,\\*[0.2cm]
\frac{\partial L}{\partial v^A}\simeq\frac{k+1}{2k+1}m_kv^{-2k/(2k+1)}\ast
v_A,&
\frac{\partial L}{\partial q^A}\simeq -(l+1)cq^{2l}\ast q_A,\\*[0.2cm]
\frac{\partial S}{\partial p_A}\simeq\frac{k+1}{m}p^{2k}\ast p_A,&
\frac{\partial S}{\partial
f_A}\simeq-\frac{1}{c_l}\frac{l+1}{2l+1}f^{-2l/(2l+1)}\ast f^A,\\*[0.2cm]
\frac{\partial T}{\partial v^A}\simeq m_k\frac{k+1}{2k+1}v^{-2k/(2k+1)}\ast
v_A,
\frac{\partial T}{\partial
f_A}\simeq-\frac{1}{c_l}\frac{l+1}{2l+1}f^{-2l/(2l+1)}\ast f^A.
\end{array}$$
Examples $$2l=\left\{\ba{ll}\phantom{-}0,&\mbox{harmonic oscilator}\\
-3,&\mbox{Kepler problem}.
\ea\right.$$

\subsection*{Second order field equations}
In this section for simplicity we drop the indices. Let $a$ and $b$ be a scalar
functions responsible for the nonlinear theories. The results of the previous
section suggest to consider the closed ideals generated by the relations
\begin{eqnarray}
\{(q,p)-\mbox{subbundle}&|&dq-\ast e^bp,dp-\ast
e^aq,\delta(e^aq),\delta(e^bp)\}
\{(q,v)-\mbox{subbundle}&|&dq-v,\delta(e^{-b}v)-e^aq,dv,\delta(e^aq)\}\nonumber\
\{(f,p)-\mbox{subbundle}&|&dp-f,\delta(e^{-a}f)-e^bp,df,\delta(e^bp)\}\nonumber\
\{(f,v)-\mbox{subbundle}&|&d(\ast e^{-a}f)-v,d(\ast e^{-b}v)-f,dv,df\}
\end{eqnarray}
In the notations of the electromagnetic field theory, gauge potential
$q\rightarrow A,$ field strength $v\rightarrow F,$ induction $p\rightarrow G$
and current $f\rightarrow j,$ the Maxwell's equations on four different
Poincar\'e-Cartan subbundles have the forms
$$\begin{array}{cc}
\left\{\begin{array}{c}dA=F\\\delta(e^{-b}F)=e^aA\end{array}\right.&
\left\{\begin{array}{c}dA=\ast e^bG\\dG=\ast e^aA\end{array}\right.\\*[0.7cm]
\left\{\begin{array}{c}\delta(e^{-a}j)=e^bG\\dG=j\end{array}\right.&
\left\{\begin{array}{c}\delta(e^{-a}j)=\ast F\\\delta(e^{-b}F)=\ast
j.\end{array
\end{array}$$

If $|f|=0,$ then $\delta(f\alpha)=f\delta\alpha+i_{gdf}\alpha.$
Using this one can equivalently rewrite the above equations in the Dirac form
as
was noticed by Marcel Riesz in 1946, by means of the Dirac operator $D\equiv
d+\delta.$ For example
$$\left\{\begin{array}{c}DA=F-i_{gda}A,\\DF=e^{a+b}A+i_{gdb}F.\end{array}\right.

The above first order relations lead to the second order relations (field
equations) of {\em two} types,
$$\begin{array}{lcl}
\delta dq=e^{a+b}q+i_{g{\rm d}b}dq&\mbox{and}&\delta q=-i_{g{\rm d}a}q;\\
\delta dp=e^{a+b}p+i_{g{\rm d}a}dp&\mbox{and}&\delta p=-i_{g{\rm d}b}p;\\
d\delta
f=e^{a+b}f+{\rm d}(a+b)\wedge(\delta-i_{g{\rm d}a})f+di_{g{\rm
d}a}f&\mbox{and}&
d\delta
v=e^{a+b}v+{\rm d}(a+b)\wedge(\delta-i_{g{\rm d}b})v+di_{g{\rm
d}b}v&\mbox{and}&
\end{array}$$
\noindent Therefore
\be \triangle q=e^{a+b}q+i_{g{\rm d}b}dq-di_{g{\rm d}a}q,\ee
and the same for $\triangle p$ if we interchange $q\leftrightarrow p$ and
$a\leftrightarrow b$ in (38).
For the ``harmonic oscilator'' case, $a$ and $b$ are constants, the second
order
equations for $q$ and $p$ are the London
equations. In case of mechanics, strings and electrostatics, $|q|=0,$
the ``gauge fixing'' condition, $\delta q=0,$ is an
identity and the equation (38) is the Newton equation.

A scalar function $a$ is responsible for the nonlinear relation among current
and gauge potential, $j=\ast aA.$ The conservation of current , $dj=0,$ is
equivalent to the $a$-{\em dependent} gauge condition
$\delta(aA)=0\;\Rightarrow\;\delta A=-i_{g{\rm d}a}A.$ The Lorentz gauge,
$\delta A=0,$ is possible only if $a=$const.

\subsection*{References}
\begin{description}
\item Gaw{\c e}dzki Krzysztof (1972), Reports on Mathematical Physics {\bf 3}
(4) 307-326
\item Gusiew Magdalena (1993), {\em  Formalizm multisymplektyczny w mechanice i
elektrodynamice klasycznej}, Diploma Thesis, University of
Wroc{\l}aw,\\Institute of Theoretical Physics
\item Henneaux Marc and Claudio Teitelboim (1986), {\em p-Form
Electrodynamics},
Foundations of Physics {\bf 16} (7) 593-617
\item Jadczyk Arkadiusz and Marco Modugno (1992), {\em New geometrical approach
to Galilei general relativistic quantum mechanics}, XXI Conference on
Differential Geometry Methods in Theoretical Physics, China
\item Kijowski Jerzy (1973), {\em A finite-dimensional canonical formalism in
the classical field theory}, Communications in Mathematical Physics {\bf 30}
99-128
\item Kijowski Jerzy and W{\l}odzimierz Tulczyjew (1979), {\em A symplectic
framework for field theories}, Lectures Notes in Physics, vol {\bf 107},
Springer-Verlag, Berlin.
\item Kondracki Witold (1978), Reports on Mathematical Physics {\bf 16} (1)
9-47
\item Krupkova Olga (1992), {\em Higher order mechanics}, Habilitation
Thesis,\\
University of Opava
\item Oziewicz Zbigniew and Wojciech Gruhn (1983), {\em On Jacobi's theorem
from
a year 1838}, Hadronic Journal {\bf 6} (6) 1579-1605
\item Oziewicz Zbigniew (1986), {\em From Grassmann to Cliford}, in {\em
Clifford Algebras and their Applications in Mathematical Physics}, edited by
J.S.R. Chisholm and A.K. Common, R. Reidel Publishing Company,\\ Dordrecht,
NATO
ASI C-183, 245-255
\item Oziewicz Zbigniew (1992), {\em Calculus of variations for multiple-valued
functionals}, Reports on Mathematical Physics {\bf 31} (1) 85-90
\item Piron Constantin (1983), {\em New quantum mechanics}\\ In {\em Old and
New
Questions in Physics, Cosmology, Philosophy and Theoretical Biology. Essays in
Honour of Wolfgang Yourgrau}, edited by Alwyn Van Der Merve. Plenum Press, New
York and London.
\item Piron Constantin (1989), {\em Lectures on Electrodynamics},
l'Universit\'e de Gen\`eve
\item Piron Constantin (1990), {\em M\'ecanique Quantique: Bases et
Applications}, Presses polytechniques et universitaires romandes Lausanne
\item Piron Constantin (1992), {\em Theorie de la Gravitation d'Einstein},\\
l'Universit\'e de Gen\`eve
\item Thirring Walter (1979), {\em Classical Field Theory}. Springer,\\Berlin,
Heidelberg, New York
\item Tulczyjew W. M. (1979), Reports on Mathematical Physics {\bf 16} 233
\end{description}
\end{document}